\documentclass[reprint,reprintnumbers,amsmath,amssymb]{revtex4-1}

\usepackage{graphicx}
\usepackage{dcolumn}
\usepackage{bm}

\usepackage{epstopdf}
\usepackage[flushleft]{threeparttable}

\usepackage{siunitx}

\newcommand{\s}{\mathrm{S}}
\newcommand{\g}{\mathrm{G}}
\newcommand{\T}{\mathrm{T}}
\newcommand{\aIGZO}{\textit{a}-IGZO }

\begin{document}

\title{Universal dependence on the channel conductivity of the competing weak localization and antilocalization in amorphous InGaZnO$_4$ thin-film transistors}

\author{Wei-Hsiang Wang}
\altaffiliation{These authors contributed equally to this work.}
\affiliation{Department of Physics, National Taiwan Normal University, Taipei 116, Taiwan}
\author{Syue-Ru Lyu}
\altaffiliation{These authors contributed equally to this work.}
\affiliation{Department of Physics, National Taiwan Normal University, Taipei 116, Taiwan}
\author{Elica Heredia}
\affiliation{Department of Physics, National Taiwan Normal University, Taipei 116, Taiwan}
\author{Shu-Hao Liu}
\affiliation{Department of Physics, National Taiwan Normal University, Taipei 116, Taiwan}
\author{Pei-hsun Jiang}
\altaffiliation{E-mail: pjiang@ntnu.edu.tw}
\affiliation{Department of Physics, National Taiwan Normal University, Taipei 116, Taiwan}
\author{Po-Yung Liao}
\affiliation{Department of Physics, National Sun Yat-Sen University, Kaohsiung 804, Taiwan}
\author{Ting-Chang Chang}
\affiliation{Department of Physics, National Sun Yat-Sen University, Kaohsiung 804, Taiwan}

\begin{abstract}
We investigate the gate-voltage dependence of the magnetoconductivity of several amorphous InGaZnO$_4$ (\textit{a}-IGZO) thin-film transistors  (TFTs). The magnetoconductivity exhibits gate-voltage-controlled competitions between weak localization (WL) and weak antilocalization (WAL), and the respective weights of WL and WAL contributions demonstrate an intriguing universal dependence on the channel conductivity regardless of the difference in the electrical characteristics of the \aIGZO TFTs. Our findings help build a theoretical interpretation of the competing WL and WAL observed in the electron systems in \aIGZO TFTs.
\end{abstract}

\maketitle

Extensive use of integrated circuits with low power consumption \cite{Rodbell2007} is required in the fabrication of modern electronic devices. Among these consumer products, thin-film-transistor (TFT) nonvolatile memory devices based on oxide semiconductors have recently attracted great attention owing to their potential application in flexible system-on-panel displays \cite{Yin2008,Chen2010,Seo2009,Lee2009}. Amorphous InGaZnO$_4$ (\textit{a}-IGZO) TFTs, in particular, have several advantages over other transparent conducting oxides, such as high field-effect mobilities \cite{Nomura2004, Na2008,Kim2007,Han2009,Kamiya2009,Suresh2008}, small subthreshold swings with low off-state currents \cite{Su2011},  good uniformity, and tunable carrier concentrations even when deposited at room temperature \cite{Hosono2006,Nomura2006}. Therefore, they are a promising alternative to amorphous silicon TFTs as switching and driving devices for application in active-matrix liquid-crystal displays and organic light-emitting diode displays.

Apart from the investigation of a few quality issues of \aIGZO TFTs for practical applications conducted so far \cite{Chen2011,Kang2007,Park2008},  research on their fundamental electrical properties at low temperatures is necessary for exploring quantum corrections to the conductivity of these carrier systems with disorders. Quantum interference and weak localization have been studied in electron systems in various materials \cite{Hikami1980,Altshuler1980,Ando1982,Bergmann1984,Abrahams2001,Lee1985,Lin2002,Pierre2003}  
.  InZnO semiconductor films and nanowires, in particular, have raised special interest because of their intricate physical properties and potential applications in nanoelectronic and spintronic devices \cite{Shinozaki2007,Shinozaki2013,Yabuta2014,Thompson2009,Chiu2013,Kulbachinskii2015,Xu2010,Ozgur2005}.  However, few comprehensive in-depth studies of the low-temperature electrical transports in InZnO and similar multicomponent oxide semiconductors were condected. Recently, we reported an intriguing observation of controllable competition between weak localization (WL) and weak antilocalization (WAL) in the electron systems in \aIGZO TFTs under variation of the gate voltage or the temperature \cite{Wang2017},  but the underlying mechanism remains undetermined. More studies are required to further understand the quantum interference in these systems.

In this work, we extend our study of the competing WL and WAL observed in \aIGZO TFTs by examining the low-temperature channel magnetoconductivity (MC) of a series of \aIGZO TFTs with various channel dimensions. Manipulated via electric gating, the MC of all the samples reveals interesting competitions between WL and WAL at small magnetic fields, where the WAL component grows drastically with increasing gate voltage and the WL component stays steady but small. Further analyses of the WL and WAL components for all samples with different electrical characteristics reveal an intriguing universal dependence on the channel conductivity. The finding of this universal behavior helps provide a theoretical interpretation of the competing WL and WAL observed in the two-dimensional (2D) carrier systems in \aIGZO TFTs.

The \aIGZO TFTs based on an etch-stopper structure with passivation capping were fabricated on glass substrates. 300-nm-thick Ti/Al/Ti was first deposited as the gate electrodes by radio-frequency magnetic sputtering, followed by a 300-nm-thick SiN$_x$ gate insulating layer using plasma-enhanced chemical vapor deposition. Next, 60-nm-thick \aIGZO was deposited by sputtering at room temperature to serve as the channel layer, followed by sputter deposition of a 200-nm-thick organic material as the etching stop layer, and then 300-nm-thick Ti/Al/Ti as the source and drain electrodes. The fabrication was finished with sputter deposition of 200-nm-thick organic material as the passivation layer. The geometries of the channel and the electrodes were patterned using standard photolithography techniques.

\begin{figure}
\centering
\includegraphics[width=2.1in]{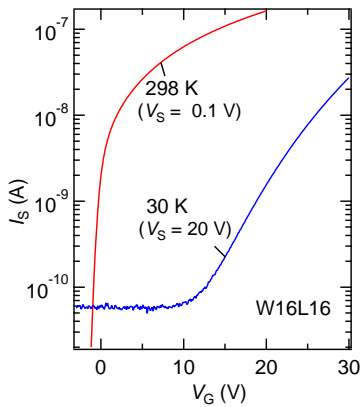}
\caption{
Source currents ($I_\s$) of a representative \aIGZO TFT (W16L16) as functions of the gate voltage ($V_\g$) at room temperature with $V_\s=0.1$ V and at $T=30$ K with $V_\s=20$ V, where $V_\s$ is the source-to-drain voltage.}
\label{fig:fig1}
\end{figure}

In our experiment, we performed electrical measurements by monitoring the source-to-drain currents ($I_\s$) when applying source-to-drain voltages ($V_\s$) and gate-to-drain voltages ($V_\g$) to several \aIGZO TFTs with channel lengths ($L$) and widths ($W$) ranging from 2 to 16 $\mu$m. The room-temperature measurements on our TFT samples show that the threshold gate voltages ($V_\T$) lie between $-4.5$ and 0.9 V, the subthreshold swings (SS $= dV_\g/d(\log I_\s$)) lie between 0.3 and 1.3 V/decade, and the mobilities lie between 4.2 and 17 cm$^2$/Vs. These characteristics differ moderately and somewhat randomly from sample to sample without direct relation to the variation of the channel dimensions. As each sample is cooled to cryogenic temperatures ($T$),  the mobility decreases, while $V_\T$ and SS increase, with  $V_\T$ starting to show a tendency to increase with $L$ for any fixed $W$. At the base temperature, which is 1.5 K for our cryogenic system, samples with large $L/W$ get ``frozen'', making  $I_\s$ undetectable even under considerable $V_\g$ and $V_\s$. Therefore, as one of the control variables, a fixed $T$ of 30 K has been chosen for the following experiment to obtain $I_\s$ with reasonable signal-to-noise ratios for all the samples, while the electron systems in \aIGZO TFTs are still kept in the quantum diffusive regime \cite{Wang2017, Shinozaki2013}.  Typical $I_\s$--$V_\g$ characteristics of an \aIGZO TFT at room temperature and at 30 K can be seen from a representative sample shown in Fig.~\ref{fig:fig1}.

\begin{table}[]
\centering
\caption{Channel dimensions and low-temperature (30-K) electrical characterization of the representative \aIGZO TFT samples.}
\label{tab:tab1}
\begin{center}
{\renewcommand{\arraystretch}{1.2}
\begin{tabular}{lccccc}
\hline\hline
{Sample} & \begin{tabular}{@{}c@{}}$W$ \\ ($\mu$m)\end{tabular} & \begin{tabular}{@{}c@{}}$L$ \\ ($\mu$m)\end{tabular} & \begin{tabular}{@{}c@{}}$V_\T$ \\ (V)\end{tabular} & \begin{tabular}{@{}c@{}}SS \\ {(V/decade) }\end{tabular} & \begin{tabular}{@{}c@{}}$\sigma_{22\mathrm{V}}$\tnote{a}$^\mathrm{a}$ \\ ($10^{\text{-}6} e^2/h$)\end{tabular} \\ \hline
W16L4 & \multicolumn{1}{S[table-format=3.2]}{16} & \multicolumn{1}{S[table-format=3.2]}{4} & \multicolumn{1}{S[table-format=3.2]}{-2.0} & \multicolumn{1}{S[table-format=3.2]}{3.1} & 127 \\
W16L8 & \multicolumn{1}{S[table-format=3.2]}{16} & \multicolumn{1}{S[table-format=3.2]}{8} & \multicolumn{1}{S[table-format=3.2]}{2.9} & \multicolumn{1}{S[table-format=3.2]}{7.2} & 48.6 \\
W16L16 & \multicolumn{1}{S[table-format=3.2]}{16} & \multicolumn{1}{S[table-format=3.2]}{16} & \multicolumn{1}{S[table-format=3.2]}{11.4} & \multicolumn{1}{S[table-format=3.2]}{6.1} & 72.1 \\
W2L12 & \multicolumn{1}{S[table-format=3.2]}{2} & \multicolumn{1}{S[table-format=3.2]}{12} & \multicolumn{1}{S[table-format=3.2]}{11.0\tnote{b}$^\mathrm{b}$} & \multicolumn{1}{S[table-format=3.2]}{4.3\tnote{b}$^\mathrm{b}$} & 205\tnote{b}$^\mathrm{b}$\\ \hline\hline
\end{tabular}
}
\end{center}

\begin{tablenotes}
\footnotesize\item[a]$^\mathrm{a}$ As an example to demonstrate the difference among the transport properties of the samples, $\sigma_{22\mathrm{V}}$ is the 2D channel conductivity at zero magnetic field at a moderate fixed $(V_\g-V_\T)$ value of 22 V.
\item[b]$^\mathrm{b}$ $V_\T$, SS, and $\sigma_{22\mathrm{V}}$ in this table are measured at $V_\s=20$ V, except those of W2L12, which are measured at $V_\s=40$ V.
\end{tablenotes}
\end{table}

\begin{figure} 
\includegraphics[width=3.5in]{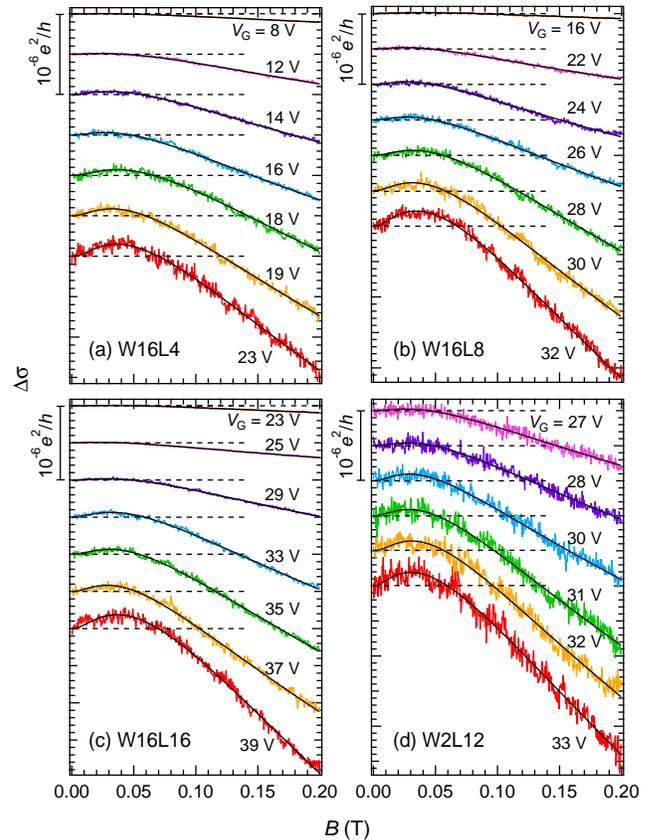}
\caption{
Normalized magnetoconductivity ($\Delta \sigma$) of representative \aIGZO TFTs as functions of the magnetic field ($B$) at various fixed gate voltages ($V_\g$) at $T=30$ K. Samples (a) W16L4, (b) W16L8, and (c) W16L16 are operated at $V_\s = 20$ V, while (d) W2L12  is operated at $V_\s = 40$ V. Traces are vertically offset for clarity with horizontal dashed lines indicating the corresponding zeros. Theoretical fits (Eq.~\ref{eq:1}) are shown as solid smooth lines.}
\label{fig:fig2}
\end{figure}

The quantum magnetotransport properties of the \aIGZO TFTs are investigated at $T=30$ K in the presence of magnetic fields ($B$) perpendicular to the channel plane. Basic information of the representative samples to be presented in this paper is listed in Table \ref{tab:tab1}. Shown in Figs.~\ref{fig:fig2}a--\ref{fig:fig2}c are the normalized MC traces ($\Delta \sigma(B)=\sigma(B)-\sigma(0)$) as functions of $B$ at various $V_\g$ with a fixed $V_\s$ of 20 V for samples W16L4, W16L8, and W16L16, respectively. W2L12 in Fig.~\ref{fig:fig2}d, however, uses $V_\s=40$ V instead for its long narrow channel to achieve $I_\s$ with acceptable signal-to-noise ratios.  Here, the conductivity $\sigma$ is defined as $I_\s/V_\s\cdot L/W$, i.e., the 2D conductivity of the channel. The traces are vertically offset for clarity with horizontal dashed lines indicating the corresponding zeros. All the $\Delta \sigma(B)$ traces are symmetric about $B=0$, but only the portions under positive $B$ are shown for simplicity. These samples exhibit similar interesting features of $\Delta \sigma(B)$ as $V_\g$ is varied. At the highest $V_\g$ applied to each sample, $\Delta \sigma(B)$ slightly increases with $B$ up to $B \sim 0.04$ T, demonstrating a WL signature. At higher $B$, however, the trace curve then bends over to WAL and decreases substantially. If $V_\g$ is decreased, the $\Delta \sigma(B)$ traces of all the samples undergo similar transitions---the maximum of $\Delta \sigma(B)$ is gradually suppressed and then smeared into a plateau below $B\sim0.05$ T at a smaller $V_\g$, while the descent of $\Delta \sigma(B)$ under higher $B$ remains significant. When $V_\g$ drops further, approaching $V_\T$, the $B$ dependence of the entire $\Delta \sigma$ trace fades out, exhibiting a behavior like a \textit{unitary} state \cite{Hikami1980}. These behaviors imply that, in \aIGZO TFTs, competitions between WL and WAL occur and can be controlled by $V_\g$.

\begin{figure}
\includegraphics[width=3.6in]{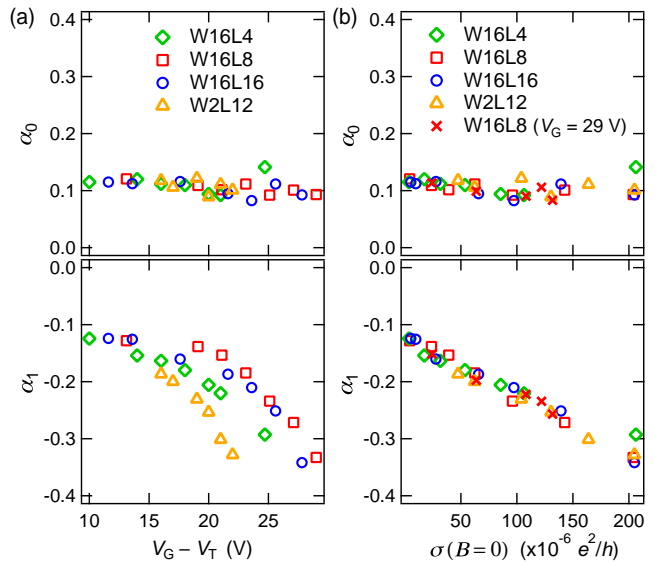}
\caption{\label{fig:fig3}
$\alpha_0$ (the prefactor for WL) and $\alpha_1$ (the prefactor for WAL) obtained from the theoretical fits in Fig.~\ref{fig:fig2} as functions of (a) $V_\g - V_\T$, and (b) the channel conductivity at $B=0$ ($\sigma (B=0)$), except for the crosses in (b), which is obtained from the analysis of another measurement on W16L8 at a fixed $V_\g$ of 29 V with $V_\s$ varied from 6 to 26 V. $V_\T$ is the threshold gate voltage of each respective sample.}
\end{figure}

To analyze the respective contributions of WL and WAL for each $\Delta \sigma(B)$ trace, we use the two-component Hikami--Larkin--Nagaoka (HLN) theory for the MC of a 2D system \cite{Hikami1980,Lu2011,Lu2011a}: 
\begin{equation}\label{eq:1}
\Delta\sigma(B)=A\sum_{i=0, 1}\frac{\alpha_i e^2}{\pi h}\Bigg[\Psi \Bigg(\frac{\ell_B^2}{\ell_{\phi i}^2}+\frac{1}{2} \Bigg) - \ln \Bigg( \frac{\ell_B^2}{\ell_{\phi i}^2} \Bigg)\Bigg],
\end{equation}
where $\Psi$ is the digamma function, $\ell_B \equiv \sqrt{\hbar/(e|B|)}/2$ is half the magnetic length, the prefactors $\alpha_0$ and $\alpha_1$ stand for the weights of WL and WAL contributions, respectively, and $\ell_{\phi i}$ is the corresponding phase coherence length. $A$ is a small coefficient of $10^{-4}$, which we have added to the original two-component HLN equation  to adequately present the nontrivial information retrieved from the seriously suppressed conduction of \aIGZO TFTs at low temperatures \cite{Wang2017}. Eq.~\ref{eq:1} provides excellent fits to the data, which are displayed as solid smooth lines in Fig.~\ref{fig:fig2}. $\alpha_0$ and $\alpha_1$ obtained from the fits for all the samples are summarized in Fig.~\ref{fig:fig3}a as functions of ($V_\g - V_\T$), where $V_\T$ is the threshold gate voltage of each respective sample. $\ell_{\phi i}$ values deduced from the fittings are on the order of $\sim$$10^{-1}$ $\mu$m, which is considerably smaller than the channel dimensions. 

Examination of Fig.~\ref{fig:fig3}a shows that all the samples share similar evolutions of $\alpha_0$ and $\alpha_1$ when within their respective ranges of operating $V_\g$. In these operating ranges, $|\alpha_1|$ (the prefactors for WAL) drop considerably from $\sim$0.3 to $\sim$0.1 as $V_\g$ is decreased,  whereas $\alpha_0$ (the prefactors for WL) stay near 0.1. This indicates a drastically decreasing WAL component accompanied by a small but relatively steady WL contribution for each sample as $V_\g$ is decreased. When a sample approaches its unitary-like state, $\alpha_1$ becomes close to $-\alpha_0$, leading to a flattened trace of $\Delta \sigma$ vs. $B$.

The magnitudes of $I_\s$ of the samples in this experiment bear considerable diversity, with W2L12 giving only $I_\s=5.2\times10^{-8}$ A at $V_\g = 33$ V and $V_\s=40$ V, compared to W16L4's  $I_\s=6.3\times10^{-7}$ A at $V_\g = 23$ V and $V_\s=20$ V. The basic electrical characteristics also vary from sample to sample as shown in Table \ref{tab:tab1}. Therefore, it is not surprising to find in Fig.~\ref{fig:fig3}a that the value of $\alpha_1$ of each sample evolves with ($V_\g - V_\T$) at its own different pace. (The difference in $\alpha_0$ may not be observable because their values inherently stay  $\sim$0.1 for our samples.) However, if $\alpha_0$ and $\alpha_1$ are plotted against the channel conductivity at zero magnetic field ($\sigma(B=0)$), as illustrated in Fig.~\ref{fig:fig3}b, each of them collapses onto a universal linear curve, indicating that the competition between WL and WAL is determined by some variable that also controls the zero-magnetic-field channel conductivity.
Samples other than the four representatives, though not shown in this paper, also have their $\alpha_i$ vs. $\sigma(B=0)$ curves coincide with the universal one.
Variations in other electrical characteristics, including $V_\T$, SS, and the mobility, do not affect the respective weights of WL and WAL contributions, at least within the scopes studied in this experiment. Moreover, alterations in $V_\s$ also manifest this universal behavior. The crosses displayed in Fig.~\ref{fig:fig3}b are obtained from the analysis of another representative measurement on W16L8 at a fixed $V_\g$ of 29 V with $V_\s$ varied from 6 to 26 V, leading to a moderate rise in $\sigma(B=0)$ until $\sigma(B=0)$ saturates at an even higher $V_\s$. 
This set of data also falls onto the universal curve. The intriguing universal dependence on the channel conductivity of the WL and WAL contributions indicates a certain correlation between the quantum interference and the overall transport performance. This should provide important information for future theoretical research.

The gate voltage of the \aIGZO TFT is successfully utilized in our experiment to control the competition between WL and WAL. WL and WAL behaviors are also studied for \aIGZO films \cite{Shinozaki2013,Yabuta2014}, and a crossover from WL to WAL has been found in samples with stronger spin--orbit scattering as $T$ is decreased to sufficiently low temperatures. With electric gating, our \aIGZO TFT devices provide a feasible option other than temperature variation to manipulate the quantum interference in \textit{a}-IGZO. It is noted that the WAL behaviors in our devices survive at higher $T$ ($>40$ K) \cite{Wang2017} compared to those observed in some \aIGZO films \cite{Shinozaki2013,Yabuta2014}. This is probably due to stronger spin--orbit disorder in our samples. 

Gate-voltage-controlled competitions between WL and WAL are also found in the surface states of topological insulators \cite{Lang2013}. The theoretical interpretation of those observations is that $\alpha_0$ and $\alpha_1$ are determined by the ratio of the gap opening at the Dirac point to the Fermi energy \cite{Lu2011, Lu2011a}, which can be manipulated via electric gating. In the attempt to construct an explanation for the competing WL and WAL observed in  \aIGZO TFTs, it seems plausible to associate $\sigma(B=0)$ with the Fermi-level position relative to the band structure, with a reminder that the gate bias on a TFT has an effect of band bending. Therefore, the universal dependence on $\sigma(B=0)$ of the WL and WAL components observed for \aIGZO TFTs may suggest a similar theoretical interpretation for the gate-voltage-controlled competitions. However, the mechanism that induces the crossover between WL and WAL remains unclear. Besides, with the quantum interference correction to $\sigma(B=0)$ coming into play, theoretical calculations are needed to confirm the correlations among $\sigma(B=0)$, the relative Fermi-level position, and the weights of WL and WAL for \aIGZO TFTs. More theoretical studies are required to fully understand the underlying physics.

In summary, we have measured the MC of several \aIGZO TFTs with various electrical characteristics at a cryogenic temperature of 30 K. The MC of each sample at a large ($V_\g-V_\T$) slightly increases with $B$ up to $B \sim 0.04$ T, exhibiting a WL behavior, but then bends over to WAL and decreases substantially at higher $B$. The maximum of the MC curve is gradually suppressed into a plateau below $B\sim0.05$ T with decreasing $V_\g$, before the entire trace of MC as a function of $B$ is flattened into the unitary class at an even lower $V_\g$.   A data  analysis using the two-component HLN theory reveals a gate-voltage-controlled competition between WL and WAL, where the WL component remains near a small value, while the WAL component grows drastically with increasing $V_\g$. Further examination of the WL and WAL components  for all samples reveals an intriguing universal dependence on the zero-magnetic-field channel conductivity. This may imply  that the gate-voltage-controlled competition between WL and WAL can be interpreted in terms of the quantum interference being tuned by the Fermi-level position relative to the band structure. However, more theoretical studies are necessary to fully understand the underlying physics of the competing WL and WAL in the carrier systems hosted in \aIGZO TFTs.

The work was supported by the Ministry of Science and Technology of the Republic of China under Grant No. MOST 102-2112-M-003-009-MY3.

\end{document}